\documentclass[apj]{emulateapj}
\usepackage{graphicx,epsfig}

\newcommand{\nii}{[N\thinspace{II}]}
\newcommand{\oii}{[O\thinspace{II}]}
\newcommand{\oiii}{[O\thinspace{III}]}

\begin{document}
\ifx\href\undefined\else\hypersetup{linktocpage=true}\fi
\title{The Luminosity Function of L\lowercase{y}$\alpha$ Emitters at Redshift
     \lowercase{$z$}$\sim5.7$\altaffilmark{1}}
\author{Esther~M.\ Hu,\altaffilmark{2,3,4} 
Lennox~L.\ Cowie,\altaffilmark{2,3,4} 
Peter~Capak,\altaffilmark{2,3,4} 
Richard~G.\ McMahon,\altaffilmark{5} 
Tomoki~Hayashino,\altaffilmark{6} 
and
Yutaka~Komiyama\altaffilmark{7} 
}

\altaffiltext{1}{Based in part on data collected at the Subaru Telescope,
  which is operated by the National Astronomical Society of Japan}
\altaffiltext{2}{Visiting Astronomer, W. M. Keck Observatory, which is jointly
  operated by the California Institute of Technology, the University of
  California, and the National Aeronautics and Space Administration}
\altaffiltext{3}{Visiting Astronomer, Canada-France-Hawaii Telescope (CFHT), 
  which is operated by the National Research Council of Canada, the Centre
  National de la Recherche Scientifique of France, and the University of
  Hawaii.}
\altaffiltext{4}{Institute for Astronomy, University of Hawaii,
  2680 Woodlawn Drive, Honolulu, HI 96822; hu@ifa.hawaii.edu; 
  cowie@ifa.hawaii.edu; capak@ifa.hawaii.edu}
\altaffiltext{5}{Institute of Astronomy, Madingley Road,
  Cambridge CB3 OHA, U.K.; rgm@ast.cam.ac.uk}
\altaffiltext{6}{Research Center for Neutrino Science, Graduate
  School of Science, Tohoku University, Sendai 980-8578, Japan;
  haya@syssrv.awa.tohoku.ac.jp}
\altaffiltext{7}{Subaru Telescope, National Astronomical
  Observatory of Japan, 650 N. A'ohoku Place, Hilo, HI 96720;
  komiyama@subaru.naoj.org}

\shorttitle{LUMINOSITY FUNCTION OF REDSHIFT $z\sim5.7$ LYMAN ALPHA EMITTERS}
\shortauthors{Hu et al.\/}

\slugcomment{Revised Version for Publication in the Astronomical Journal}

\begin{abstract}

  We report the results of a wide-field narrowband survey for redshift
  $z\sim5.7$ Ly$\alpha$ emitters carried out with the $34' \times 27'$
  field-of-view SuprimeCam mosaic CCD camera on the Subaru 8.3-m
  telescope. Deep narrowband imaging of the SSA22 field through a 120 \AA\
  bandpass filter centered at a nominal wavelength of 8150 \AA\ was 
  combined with deep multicolor $RIz'$ broadband imaging with SuprimeCam, 
  then supplemented with $BVRZ$ imaging taken with the $42' \times 28'$ 
  field-of-view CFH12K camera on the Canada-France-Hawaii 3.6-m telescope 
  to select high-redshift galaxy candidates.  Spectroscopic observations 
  were made using the new wide-field multi-object DEIMOS spectrograph on 
  the 10-m Keck\,II Telescope for 22 of the 26 candidate objects. Eighteen 
  of these objects were identified as $z\sim5.7$ Lyman alpha emitters, and 
  a further nineteenth object from the candidate list was identified 
  based on an LRIS spectrum.  At the 3.3 \AA\ resolution of the DEIMOS
  observations the asymmetric profile for Ly$\alpha$ emission with its
  steep blue fall-off can be clearly seen in the spectra of the identified
  galaxies. This is by far the largest spectroscopic sample of galaxies
  at these redshifts, and we use it to describe the distribution of
  equivalent widths and continuum color break properties for identified
  Ly$\alpha$ galaxies compared with the foreground population. The
  large majority (at least 75\%) of the lines have rest frame Ly$\alpha$
  equivalent widths substantially less than 240~\AA\ and can be understood
  in terms of young star forming galaxies with a Salpeter initial mass
  function for the stars.  With the narrowband selection criteria of
  ($I-N) > 0.7$ and $N<25.05$ (AB mags) we find a surface density of
  Ly$\alpha$ emitters of 0.03 per square arcminute in the filter bandpass
  ($\Delta z=0.1$) down to a limiting
  flux of just under $2\times10^{-17}$ erg cm$^{-2}$ s$^{-1}$. The
  luminosity function of the Ly$\alpha$ emitters is similar to that
  at lower redshifts to the lowest measurable luminosity of $10^{43}$
  erg s$^{-1}$ as is the universal star formation rate based on their
  continuum properties. However, we note that the objects are highly
  structured in both their spatial and spectral properties on the angular
  scale of the fields ($\approx 60$ Mpc), and that multiple fields will
  have to be averaged to accurately measure their ensemble properties.

\end{abstract}


\section{Introduction}
\label{secintro}

The study of high-redshift galaxies in the early universe has seen
substantial progress in recent times, with the discovery of objects out to
$z=6.5$ (Hu et al. 2002, Kodaira et al. 2003) and with the rapidly increasing
number of galaxies discovered at redshifts $z\gg5$ (\citealp*{dey98};
Hu, Cowie, \& McMahon 1998; \citealp*{wey98}; Hu, McMahon, \& Cowie
1999; \citealp*{stern99,ellis,aji02,tan03,rho03,mai03,cuby03,bunk03}).
Active galaxies associated with radio galaxies \citep{wil99} and AGN
discovered in deep X-ray exposures with CHANDRA \citep{cdfn} have
been seen out to $z\sim5.19$, and at optical wavelengths quasars have
been found with the Sloan survey out to  $z=6.42$ \citep{fan01,fan03}.
However, because of the relatively sparse distribution of detectable $z >
5$ galaxies it has hitherto been difficult to confirm more than a few at
a time, and thus to develop the large samples needed to understand the
robustness of candidate selection criteria and the spatial structure of
objects on the sky.  Consequently there are large uncertainties in the
inferred luminosity function and universal star formation rates.

The availability of mosaic CCDs allows wide-field coverage of 
the sky on scales of half a degree to a degree
at a time \citep[e.g.,][]{mal02,ouchi03,aji03}.  
This gives us the opportunity to generate large samples
of objects to examine their statistical properties
and construct more precise luminosity functions.

This paper is the first in a series which will describe searches for
Ly$\alpha$ emission-line galaxies at $z=5.7$ and $z=6.5$ in a number of
such wide fields. It represents a continuation of our narrowband filter
survey \citep{cow98,smitty,la_lum,la6}, which probed for Ly$\alpha$
emission from $z=3.4\to6.5$ using smaller area surveys.  We use the
wide-field capabilities of SuprimeCam \citep{supcam} on the Subaru 8.3-m
Telescope to obtain samples of narrowband selected Ly$\alpha$ emitters.
Following \citet{smitty} we use narrowband filters corresponding to
$z\sim 5.7$ and $z\sim 6.5$ Ly$\alpha$ emission lines which lie in dark
spectral portions of the night sky. We combine these with extremely deep
continuum observations to select the  high-$z$ emission-line galaxies in
a statistically uniform fashion. We then use the new wide-field DEIMOS
spectrograph on the Keck II 10-m telescope to obtain a nearly complete
spectroscopic followup.

In the present paper we describe our survey of $z\sim 5.7$ Lyman
alpha emitters in a 700 square arcminute area around the SSA22 field
\citep{cowie_1}.  The existence of extremely deep multi-color $BVRIz'$
continuum images for this field allows a detailed examination of the
effectiveness of color selection criteria in identifying Ly$\alpha$
candidates at $z\sim5.7$, similar to our studies at $z\sim 3.4$
\citet{cow98}.  In particular, we can examine the robustness of color
thresholds for the selection of Ly$\alpha$ emitters as contrasted with
\oiii\ and \oii\ emission-line galaxies.

Discriminating Ly$\alpha$ emitters from lower $z$ emission-line objects can be
difficult at lower redshifts \citep{stern99}, but at higher
redshifts the continuum break signature becomes so extreme that there is
little likelihood of a misidentification provided we have sufficiently
deep continuum images.  
Songaila and Cowie \citeyearpar{gpnot} have used spectra of the
bright quasars to measure the average transmission of the Ly$\alpha$ forest
region as a function of redshift over the $z=4\to6$ range.  Their results
translate to a magnitude break across the Ly$\alpha$ emission line of
\begin{displaymath}
	\Delta{m} = 3.8 + 20.3\ {\rm log_{10}} \left ( {1+z}\over{7} \right )
        \label{eq:1}
\end{displaymath}
For $z=5.7$ systems this corresponds to a 3.4 magnitude break, and color
measurements alone can be used to provide a substantial selection of high-$z$
galaxy candidates 
\citep[e.g.,][]{iwa03,stan03a,bunk03,leh03,bouw03,dic03}.
However, because the continuum magnitudes
are extremely faint only the deepest observations currently
feasible can find such objects, while the emission-line objects 
are much easier to find. Furthermore, continuum selected objects
can be extremely hard to confirm spectroscopically,
while the narrowband selected objects are generally
easy to identify.  
Thus the narrowband filter samples provide an invaluable tool
for identifying large homogeneous samples of high-redshift galaxies.

With high-resolution spectroscopic followups of the candidates it is
possible to both confirm that the emission is due to red-shifted Ly$\alpha$
and to estimate the statistics of interlopers. For SSA22 our narrowband and
color selections yield an initial sample of 26 candidate $z\sim 5.7$ emitters
in the field.  The powerful new wide-field spectroscopic capabilities of the
DEIMOS spectrograph \citep{deimos03} used on the Keck 10-m telescope, with an
approximate field of $16\farcm7 \times 5\farcm0$ and good red sensitivity,
allow us to follow up these candidates in an extremely efficient manner.
We have now observed 23 of the 26 objects with either this instrument or
with the LRIS spectrograph \citep{lris} on Keck.  This results in the
confirmation of 19 new redshift $z\sim5.7$ galaxies. The 18 candidates
observed with DEIMOS  unambiguously show asymmetric profiles with the high
resolution observing mode possible with this red-optimized spectrograph.

In Section~\ref{secdata} we summarize the narrowband and 
continuum imaging together with the spectroscopic observations. 
Section~\ref{secanal} describes the color selection
and color-color plots for our selection criteria.  The
spectroscopic identifications and colors of candidates 
are tabulated and labeled in these diagrams. 
In Section~\ref{secclass} we discuss the spectroscopic
identifications of Ly$\alpha$
emitters, and the incidence and properties of foreground emitters.
In Section~\ref{secdisc} we discuss the spatial structure of the objects in
the fields and show that they are highly correlated. We compare the shapes
and equivalent widths of the lines with those of lower redshift samples and
discuss their interpretation.  Finally, the Ly$\alpha$ and UV continuum
luminosity functions of the identified $z\sim5.7$ candidates are constructed 
and compared with lower redshift
samples. We argue that the star formation rates at $z\sim5.7$ are
probably at least comparable to those at lower redshifts, but postpone a
more detailed discussion until we can use an ensemble average of a number
of wide fields.

\section{Observations}
\label{secdata}

\subsection{Optical Narrow Band-Selected Survey}
\label{secnarrow}

\begin{figure}[h]
  \centering
  \includegraphics[width=2.8in,angle=90,scale=0.9]{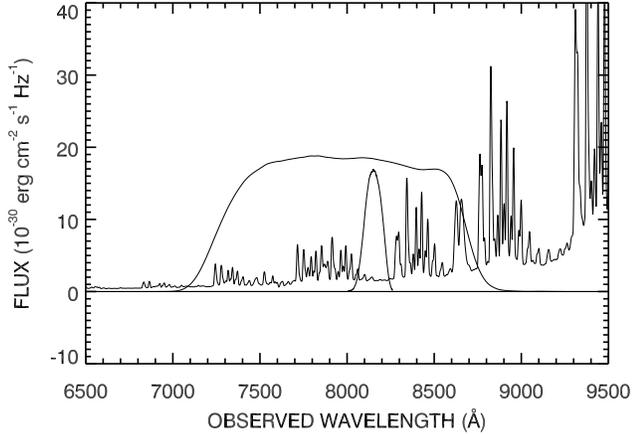}
  \caption{Plot of the NB816 and Cousins $I$ filter profiles in the f/1.86 
    beam of SuprimeCam superposed on a night-sky spectrum obtained with the
    LRIS spectrograph on Keck.  The narrowband filter is situated in a
    region of low sky background and is well centered in the $I$-band
    filter.  Due to the f/1.86 beam, the profile
    has a Gaussian shape, in contrast to the squarer profiles that would
    apply for parallel or slower beams.  Peak throughput is $\sim85\%$.
    The strength of night-sky OH airglow lines is highly variable with
    time, and the spectrum is meant to illustrate regions of high and
    low sky background rather than providing an absolute calibration.
  \label{fig1:nb816-filt}
}
\end{figure}

The present images were obtained using a 120 \AA\ (FWHM) filter centered on
a nominal wavelength of 8150 \AA\ in a region of low sky background between 
OH bands.
(The nominal specifications for the Subaru filters may be found at 
\url{http://www.naoj.org/Observing/Instruments/SCam/sensitivity.html},
and are also described in \citep{aji03}.)\ This
corresponds to Ly$\alpha$ at a redshift $z\sim 5.7$, similar to the
narrowband wavelengths probed in earlier imaging studies with the
LRIS instrument on the Keck 10-m telescopes \citep{smitty}.
The filter profile in the $f/1.86$ beam is shown in 
Figure~\ref{fig1:nb816-filt}.
The Gaussian shape of the filter profile used on SuprimeCam
may be compared with the squarer profile of the 108 \AA\ bandpass
8185 \AA\ filter used in the LRIS parallel beam for the Keck studies
\citep{hy-z}. The narrowband NB816 filter is well-centered in the
Cousins $I$ band filter, which we use as a reference
continuum bandpass for our detection of Lyman alpha emission.
Hereafter, we refer interchangeably to this filter as NB816 or
as $N$.  All narrowband and $Z$-band magnitudes are given in the AB
system.

Narrow-band images were taken with SuprimeCam on the 8.3-m Subaru
telescope on the nights of UT 2001 25--26 June and 2001 20 October
under photometric or near photometric conditions.  The imaging 
observations are summarized in Table \ref{tbl-1}, with exposures
made during times of heavy cirrus shown in parentheses.
The data were taken as a sequence of dithered background-limited exposures
and successive mosaic sequences were rotated by
90 degrees. Corresponding continuum exposures
were always obtained in the same observing run as the narrowband
exposures to avoid false identifications of transients
such as high-$z$ supernovae, or Kuiper belt objects, as Ly$\alpha$
candidates. A detailed description of the full reduction procedure 
for images is given in \citet{capak}. 
In addition to the SuprimeCam multi-color imaging on SSA22, wide-field
observations with the CFH12K camera \citep{cfh12k} were obtained 
(Table \ref{tbl-2})
for a total of 5hr 20min in $B$, 4hr 15min in $V$, 7hr 3min in $R$, and
7hr 15min in $Z$ on the CFHT 3.6-m telescope.

%
%
\begin{deluxetable}{crrc}
\tablecaption{Summary Log of SuprimeCam SSA22 Imaging\label{tbl-1}}
\tablewidth{0pt}
\tablehead{            &                   &
\colhead{$t_{\rm{exp}}$\tablenotemark{a}}  & \colhead{FWHM}  \\[0.5ex]
\colhead{Filter}       & \colhead{UT Date} &
\colhead{(s)}          &  \colhead{($''$)} }
\startdata
\noalign{\vskip-2pt}
$R$ &  20 Oct 2001 & 3000 & 0.55\\
$I$ &  24 Jun 2001 & (5040) &0.7--0.8\\
    &  26 Jun 2001 & 3300 & 0.8\\
$i'$ & 16,18 Oct 2001 & (3900) & 0.6--0.8\\
    &  22 Oct 2001 & 3300 & 0.5--0.6\\
NB816  & 25--26 Jun 2001; 20 Oct 2001 & 17700 & 0.6--0.9\\
    &  16 Oct 2001 & (12000) & 0.6\\
$z'$ & 17 Oct 2001 & (4110) & 0.7\\
    &  20 Oct 2001 & 4290 & 0.5--0.6\\
    & 10--11 Sep 2002 & 5400 & 0.5--0.7\\
\enddata
\tablenotetext{a}{Cloudy periods shown in parentheses.}
\end{deluxetable}

%
%
\begin{deluxetable}{crrc}
\tablecaption{Summary Log of CFH12K SSA22 Imaging\label{tbl-2}}
\tablewidth{280pt}
\tablehead{            &                   &
\colhead{$t_{\rm{exp}}$\tablenotemark{a}}  & \colhead{FWHM}  \\[0.5ex]
\colhead{Filter}       & \colhead{UT Date} &
\colhead{(s)}          &  \colhead{($''$)} }
\startdata
\noalign{\vskip-2pt}
$B$ & 20 Jul 2001 &  4800 & 0.92\\
    & 21 Jul 2001 &  4800 & 0.95\\
    & 16 Aug 2001 &  4800 & 1.30\\
    & 19 Aug 2001 &  4800 & 1.07\\
    &             &   5hr 20min\\
$V$ & 16 Aug 2001 &   1800 & 1.07\\
    & 20 Aug 2001 &  13500 & 0.89\\
    &             &   4hr 15min\\
$R$ & 18 Jul 2001 & 12775 & 0.79\\
    & 19 Jul 2001 &  4200 & 0.57\\
    & 20 Jul 2001 &  8400 & 0.68\\
    &             &   7hr 3min\\
$Z$ & 15 Aug 2001 &  8750 & 0.76\\
    & 18 Aug 2001 &  8620 & 0.93\\
    & 19 Aug 2001 &  8750 & 0.82\\
    &             &  7hr 15min\\
\enddata
\tablenotetext{a}{Data used for preliminary LRIS spectroscopic selection.}
\end{deluxetable}

The SuprimeCam data were calibrated using the photometric and
spectrophotometric standard, HZ4 \citep{turnshek90,oke90}, and faint Landolt 
standard stars in the SA 95-42 field \citep{landolt92}. An
astrometric solution was obtained using stars from the USNO survey.
Limiting magnitudes ($5 \sigma$ for $3''$ diameter apertures, expressed
as AB mags) are: 25.5 ($B$), 25.6 ($V$), 26.1 ($R$), 25.7 ($I$), 25.0 ($z'$),
and 25.2 ($NB816$).  

\subsection{Spectroscopy}
\label{secspec}

\begin{figure}
\includegraphics[width=2.8in,angle=90,scale=0.9]{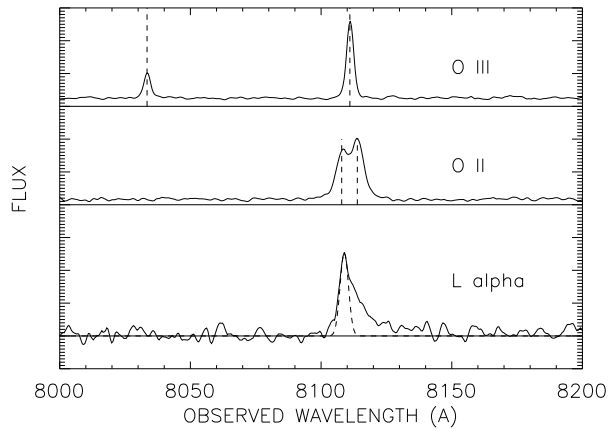}
  \caption{Composite DEIMOS spectra taken through 1$''$ wide slitlets
    with the G830 grating of [O$\,$III] (top panel), [O$\,$II] (middle panel), 
    and Ly$\alpha$ (bottom panel) emission-line objects lying within the NB816 
    bandpass.  At the 3.3 \AA\ resolution of this configuration these systems 
    are easily distinguished. Dashed vertical lines mark the positions of the 
    two components of the oxygen doublet. The stacked composite profile of the 
    Ly$\alpha$ emission systems clearly show the steep drop-off on the blue 
    side of the emission feature.  The superposed dashed profile shows the 
    instrumental response to neighboring night sky lines for comparison, and 
    indicates the blue side of the emission feature is consistent with an 
    abrupt cut-off.  The asymmetry of the profile for the Ly$\alpha$ emitters 
    is a clear signature of the forest absorption to the blue of the Ly$\alpha$
    emission line.
  \label{fig2:composite-spec}
}
\end{figure}

We used the new wide-field DEIMOS spectrograph on the Keck 2 10-m
telescope for spectroscopic followups of candidate $z\sim 5.7$
Ly$\alpha$ emitters. The G830 $\ell$/mm grating blazed at 8640
\AA\ was used with $1''$ wide slitlets.  In this configuration,
the resolution is 3.3 \AA, which is sufficient to distinguish the
$z\sim1.19$ \oii\ doublet structure from the profile of redshifted
Ly$\alpha$ emission (Fig.~\ref{fig2:composite-spec}).  Redshifted \oiii\
emitters ($z\sim0.62$), show up frequently as emission-line objects in
the narrowband and can easily be identified by the doublet signature.
The observed wavelength range ($\lambda\lambda\sim 3840$ \AA, with typical
coverage from $\sim5900-9700$ \AA) generally also encompasses redshifted
H$\beta$ and \oiii\ lines for instances where emission-line candidates
might be H$\alpha$ at $z\sim 0.24$, and deals with the problematic
instance of extragalactic \ion{H}{2} regions with strongly suppressed
\nii. The G830 grating used with the OG550 blocker gives a throughput
greater than 20\% for most of this range, and $\sim28\%$ at 8150 \AA.
The observations were taken on UT 12--13 September 2002, with the first
night showing light to moderate cirrus, and the second night clear,
and on two essentially clear nights on UT 28--29 August 2003.

Earlier spectra of a preliminary cut of candidates were also obtained
with the LRIS spectrograph during runs in UT 18--19 August 2001 and
21 October 2001. The observations were made in multislit mode using
1.4$''$ wide slits with the 400 $\ell$/mm grating blazed at 8500 \AA,
which in combination with the slit width gives a resolution of roughly
11.8 \AA.  Because of the much higher red throughput of the DEIMOS CCDs
and optics, and its wider field coverage, nearly all of the spectroscopic
identifications use the DEIMOS reduced data.

\section{Candidate Selection}
\label{secanal}

\begin{figure}
  \includegraphics[width=2.8in,angle=90,scale=0.85]{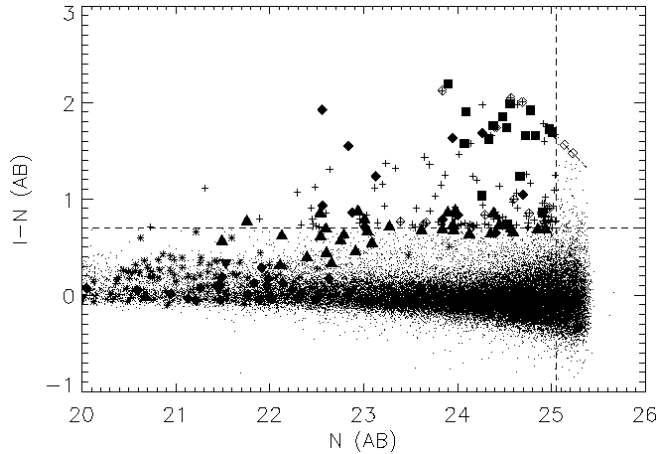}
  \caption{Plot of emission-line excess objects ($I-N$ vs. $N$) found
  in the NB816 narrowband filter as a function of object magnitude.
  Narrow-band magnitudes are given in the AB system 
  \protect{\citep{oke83,oke90}} of equivalent
  visual magnitudes.  The distribution of field objects is also shown
  as a scatter plot.  Filled symbols show the properties of spectroscopically
  identified emission-line objects (triangles are [O$\,$II] emitters,
  diamonds are [O$\,$III] emitters, and squares are candidate $z\sim5.7$
  Ly$\alpha$ emitters) and crosses inside open diamonds are candidates 
  which were spectroscopically
  observed but not identified. An $(I-N) > 0.7$ and $N < 25.05$ 
  selection criterion (dashed lines) picks out 127 objects which
  are shown with crosses if they have not been spectroscopically 
  observed. Asterisks mark the positions of known stars.
  \label{fig3:emiss-excess}
}
\end{figure}

Figure~\ref{fig3:emiss-excess} shows the $I-N_{AB}$ color as a function of
narrowband ($N_{AB}$) magnitude.  All magnitudes were
measured in $3''$ diameter apertures, and had average aperture
corrections applied to give total magnitudes.  To a
narrowband magnitude of 25.05, conservatively brighter than a 5 $\sigma$ 
cut of 25.2, there are
127 objects with $I-N_{AB}$ colors $> 0.7$ over the central
$26\farcm5 \times 26\farcm5$ field, which was conservatively chosen as a
region of uniform mosaic coverage, free of any possible edge-of-field 
illumination variation effects.  This is about $0.4\%$ of the 38451
objects with $N_{AB} < 25.05$ in this area.  As can be seen from 
Fig.~\ref{fig3:emiss-excess} a simple emission criterion chooses out a 
mixture of \oii\ and \oiii\ 
emitters in addition to the $z=5.7$
objects and a small number of stars. The emission-line excess objects have
been labeled on this plot according to their subsequent
spectroscopic identifications: filled squares for
Ly$\alpha$ emitters, triangles for \oii\ 
emitters, diamonds for \oiii\ 
emitters.

The high-$z$ objects are also expected to have very red colors because of the 
effects of the Ly$\alpha$ forest scattering and of the intrinsic Lyman 
continuum breaks in the galaxies themselves.  For a $z\sim 5.7$ object
whose Ly$\alpha$ emission is in the narrow band, the scattering in the 
$R$ band will be a mixture of Ly$\beta$ from the intergalactic medium
at $z\sim5.4$ and Ly$\alpha$ from $z\sim4.5$.  Their combined effects 
result in a transmission of only about 20\% of the light at this wavelength
based on measurements of brighter quasars (Songaila 2004, in preparation).  This
corresponds to an $R-z'_{AB}$ break of about 1.8 mags so that we expect 
the actual $R-z'$ colors for the objects to be redder than this value.  By
contrast we expect that \oii\ lines which have the Balmer and 4000 \AA\
breaks between the two bands will have weaker $R-z'_{AB}$ breaks of up to
about a magnitude, while \oiii\ or H$\alpha$ lines should have nearly
flat continua over the $R-z'$ wavelength range.  At wavelengths less than
about 6000 \AA\ we expect the galaxies to be faint owing to Lyman continuum
absorption and that this should result in a
$z\sim5.7$ galaxy being undetected in the $B$ and $V$ bands.

\begin{figure}
  \includegraphics[width=2.8in,angle=90,scale=0.9]{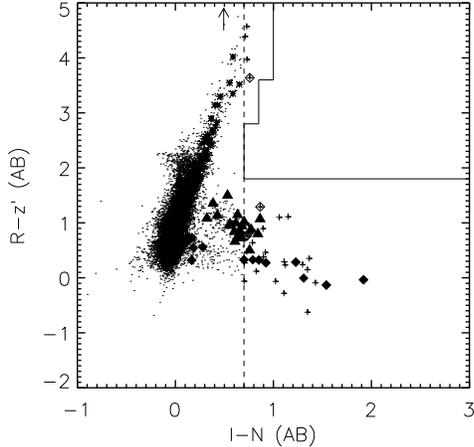}
  \caption{Plot of emission-line excess objects ($R-z'$ vs. $I-N$) found
  in the NB816 narrowband filter for a bright $N<23.8$ selection.
  The distribution of field objects is also shown
  as a scatter plot.  Filled symbols show the properties of spectroscopically
  identified emission-line objects (triangles are [O$\,$II] emitters,
  diamonds are [O$\,$III] emitters, and open diamonds
  are candidates which were spectroscopically
  observed but not identified. Plus signs indicate objects which satisfy
  the $(I-N) > 0.7$ condition (shown as the dashed line), but which have 
  not been spectroscopically observed. Asterisks show spectroscopically
  identified stars. 
  The adopted selection criteria for unambiguous $z\sim5.7$
  Ly$\alpha$ emitters in this color-color plane
  is shown by the solid lines.  This figure may be compared to a similar
  color plot (Fig.~\ref{fig6:rz-in}) for a faint narrowband selection, 
  $23.8 < N < 25.05$, which spans the magnitude range of the newly discovered
  $z\sim5.7$ Ly$\alpha$ galaxies.
  \label{fig4:rz-in-bright}
}
\end{figure}

We illustrate this in Figure~\ref{fig4:rz-in-bright} where we show the
$R-z'$ color versus the narrow band excess $I-N$ for all the brighter
objects with $N_{AB} < 23.8$ in the field.  Known \oii\ and \oiii\
emitters are marked.   At this bright narrowband magnitude cut the
color regions populated by stars and foreground galaxies are very well
defined, and we can use this to define a region (solid lines) in which
$z\sim5.7$ Lyman alpha galaxies can be unambiguously distinguished from
foreground populations.  The regions of the \oiii\ and \oii\ emitters
can clearly be seen.  As expected, the \oiii\ emitters  are nearly flat
in $R-z'_{AB}$ with typical values of 0.0 to 0.3.  These objects can
have extremely high equivalent widths, however.  By contrast the \oii\
emitters have lower equivalent widths, with only a few exceeding $I-N_{AB}
\sim 0.7$, but have redder colors.  The higher equivalent width \oii\
objects in  Figure~\ref{fig4:rz-in-bright} appear to be slighter bluer
than the weaker ones, with typical $R-z'_{AB}$ colors of just under one.
At these magnitudes there are no objects in the field with red enough
colors to be a $z\sim5.7$ emitter with $I-N_{AB} > 0.7$.

Figure~\ref{fig4:rz-in-bright} shows that red stars can lie in a simple
$I-N_{AB} > 0.7$ plus red color break selected sample since at the red
end the star track angles into this region.  The reason for this is that
the narrow band lies between strong stellar absorption features lying in
the $I$ band and these strengthen for redder stars through to L types.
This results in an apparent excess in the narrow band.  This result can be
used beneficially in color selection procedures to separate out stars from
high-$z$ galaxies or quasars (Kakazu et al.\ 2004, in preparation) but for
the present work we must also eliminate the star track.  We have therefore
used a slightly higher $I-N_{AB}$ boundary at the reddest $R-z'_{AB}$
levels which is illustrated in  Figure~\ref{fig4:rz-in-bright}.

\begin{figure}
  \includegraphics[width=2.8in,angle=90,scale=0.9]{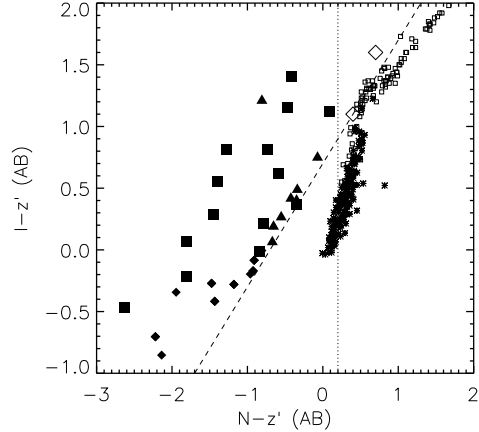}
   \caption{$I-z'$ vs. $N-z'$ color distribution of spectroscopically
  identified narrowband excess objects
  with $I-N>0.7$ (dashed line)
  compared with the spectroscopically identified stars in the field.
  Object identifications have been
  indicated with the usual plot symbols. Squares are Ly$\alpha$ emitters,  
  diamonds are [O$\,$III] emitters, triangles are [O$\,$II] emitters
  and asterisks are stars.
  The distribution of L and T dwarf stars is shown by the small open boxes
  which are drawn from the literature (Kakazu et al. 2004, in preparation)
  and lie at redder $N-z'$ colors (dotted line) compared to the identified
  Ly$\alpha$ emitters.
  Two objects in \protect{Table~\ref{tbl-3}} which are spectroscopically 
  unidentified but which are probably stars are marked with open diamonds.
  \label{fig5:iz-nz}
  } 
\end{figure}

The stars can also be distinguished from candidate Ly $\alpha$
emitters on the basis of their $N_{AB}-z'$ colors.  The location of
the star track is shown by the $I-z'$ vs $N_{AB} - z'$ color-color
plot of Fig.~\ref{fig5:iz-nz}.  The dashed line gives the $I-N_{AB} >
0.7$ selection criterion, and the stars and boxes show the track
for low-mass stars (including L and T dwarfs) for this filter system.
Because the Ly$\alpha$ emission falls within the $I$ band and can be a
substantial contributor to the measured broad-band flux, these objects
can have blue $I-z'$ colors.  In general for $I-N_{AB} > 0.7$ selected
objects the emitters lie at $N_{AB}-z' < 0.2$ and the stars lie at
$N_{AB}-z' > 0.2$. Because the level of star contamination is low and
we wished to avoid eliminating ambiguous cases we did not apply this
criterion prior to the spectroscopic observations and chose instead to
leave a small number of stars in the candidate sample. We will discuss
these objects in the section on the spectroscopy. 

\begin{figure}
  \centering
  \includegraphics[width=2.8in,angle=90,scale=0.9]{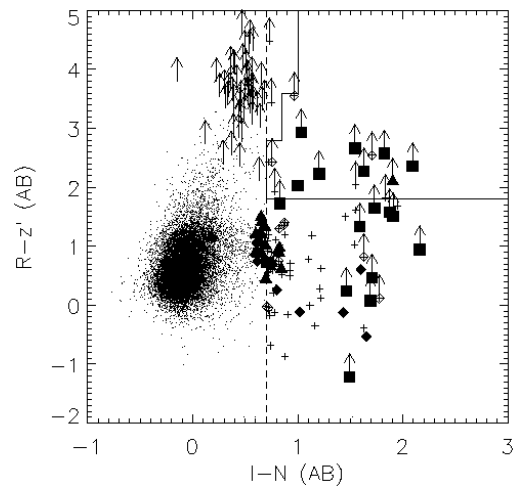}
  \caption{Color distribution of narrowband excess objects in
  $R-z'$ for the $N>23.8$ and $N<25.05$ sample.  These data may be compared
  with the bright ($N<23.8$) sample color distribution of 
  Fig.~\ref{fig4:rz-in-bright}, where the color tracks of galaxies and
  stars and the locations of foreground emission-line galaxies may
  be more clearly seen.  Object identifications have
  been indicated with the plot symbols. Squares are Ly$\alpha$ emitters, 
  triangles are  [O$\,$II] emitters, diamonds are [O$\,$III] emitters, 
  plus signs are $(I-N) > 0.7$ (dashed line) objects which were not 
  spectroscopically
  observed, and asterisks are stars.  The adopted selection
  criteria for unambiguous  $z\sim 5.7$ Ly$\alpha$ emitters indicated in
  Fig.~\ref{fig4:rz-in-bright} is shown by the solid lines.
  $R-z'$ color breaks for faint
  objects are generally limited by the $R$ band non-detection
  sensitivity.
  \label{fig6:rz-in}
  } 
\end{figure}

Our final sample of objects was then taken as all objects with
$N_{AB} < 25.05$ which clearly lay within the designated area of
$R-z'$, $I-N_{AB}$ color-color
space and which were also not detected in $V$ or $B$, together with
objects which satisfied the emission line criterion and were
not detected at any wavelength shortward of $I$.
Their $R-z'$ vs. $I-N_{AB}$ color properties are shown in
Fig.~\ref{fig6:rz-in}, 
with object symbols as previously defined.
Arrows show 2 $\sigma$ limits on the colors.  

%
%
\begin{deluxetable*}{cccccccccc}
\tablecaption{Properties of Candidate Sample\label{tbl-3}}
\tablewidth{0pt}
\tablehead{
\colhead{Object} 	& \colhead{RA(2000)} 	& 
\colhead{Dec(2000)} 	& \colhead{$N(AB)$} 	& 
\colhead{$Z(AB)$} 	& \colhead{$I$} 	& 
\colhead{$R$} 		& \colhead{$V$} 	& 
\colhead{$B$} 		& \colhead{$z$} 	}
\startdata
\noalign{\vskip-2pt}
   \phn1  &  334.220001  &  0.277625  &  25.0  &  28.7  &  28.6  &
99.0  &  99.0  &  99.0  &   5.659  \\
   \phn2  &  334.229004  &  0.093959  &  23.9  &  26.5  &  26.0  &
99.0  &  99.0  &  27.7  &   5.676  \\
   \phn3  &  334.234985  &  0.246227  &  24.5  &  24.9  &  26.3  &
28.3  &  99.0  &  99.0  &   5.660  \\
   \phn4  &  334.273010  &  0.216856  &  24.4  &  25.8  &  26.1  &
99.0  &  99.0  &  29.4  &   5.670  \\
   \phn5  &  334.278015  &  0.206272  &  24.7  &  25.2  &  26.3  &
28.3  &  99.0  &  99.0  &   5.651  \\
   \phn6  &  334.279999  &  0.462454  &  24.6  &  25.6  &  26.6  &
30.4  &  99.0  &  27.9  &  \nodata  \\
   \phn7  &  334.291992  &  0.211263  &  24.6  &  24.5  &  25.6  &
99.0  &  99.0  &  31.4  & 5.626   \\
   \phn8  &  334.308014  &  0.316606  &  24.9  &  24.5  &  25.6  &
26.9  &  99.0  &  99.0  &  \nodata  \\
   \phn9  &  334.317993  &  0.223775  &  24.6  &  25.2  &  25.8  &
28.3  &  99.0  &  29.3  &   5.645  \\
      10  &  334.320007  &  0.359441  &  25.0  &  25.5  &  25.8  &
99.0  &  99.0  &  99.0  &  \nodata  \\
      11  &  334.337006  &  0.335259  &  24.5  &  27.0  &  26.2  &
27.8  &  27.0  &  27.2  &   5.669  \\
      12  &  334.337006  &  0.293762  &  24.1  &  24.8  &  25.6  &
99.0  &  99.0  &  99.0  &   5.667  \\
      13  &  334.351990  &  0.197130  &  24.7  &  27.3  &  27.9  &
28.4  &  26.9  &  99.0  &  obs   \\
      14  &  334.369995  &  0.321620  &  24.2  &  25.0  &  25.2  &
27.1  &  28.3  &  28.2  &   5.641  \\
      15  &  334.381989  &  0.160189  &  24.5  &  25.9  &  26.5  &
28.6  &  99.0  &  27.9  &   5.684  \\
      16  &  334.388000  &  0.371196  &  24.3  &  26.1  &  25.9  &
28.0  &  26.6  &  99.0  &   5.653  \\
      17  &  334.389008  &  0.390323  &  24.6  &  23.9  &  25.5  &
27.6  &  99.0  &  26.8  &  star?   \\
      18  &  334.415985  &  0.307220  &  24.8  &  26.6  &  26.4  &
28.4  &  28.1  &  26.7  &   5.723  \\
      19  &  334.420013  &  0.404125  &  24.9  &  25.7  &  25.7  &
28.8  &  27.8  &  99.0  &   5.633  \\
      20  &  334.420013  &  0.118385  &  24.5  &  25.4  &  26.6  &
99.0  &  29.2  &  99.0  &   1.173   \\
      21  &  334.425995  &  0.128724  &  24.4  &  24.9  &  26.1  &
99.0  &  28.3  &  99.0  &   5.687\tablenotemark{a}   \\
      22  &  334.492004  &  0.228247  &  24.9  &  25.0  &  25.7  &
27.9  &  99.0  &  99.0  &  0.492\rlap{?}  \\
      23  &  334.509003  &  0.242099  &  24.1  &  25.9  &  25.9  &
99.0  &  99.0  &  99.0  &   5.673  \\
      24  &  334.519012  &  0.239439  &  24.8  &  27.4  &  28.7  &
99.0  &  99.0  &  99.0  &   5.680  \\
      25  &  334.548004  &  0.083544  &  23.8  &  25.1  &  25.9  &
28.1  &  99.0  &  28.4  &   5.709  \\
      26  &  334.556000  &  0.095670  &  25.0  &  27.2  &  26.9  &
99.0  &  99.0  &  99.0  &   5.682 \\
\enddata
\tablecomments{Magnitudes are measured in 3$''$ diameter apertures.
An entry of `99' indicates that no excess flux was measured.}
\tablenotetext{a}{Based on lower resolution LRIS data.}
\vspace*{-0.2cm}
\end{deluxetable*}

\begin{figure*}
\vspace*{-0.1cm}
\centering
\includegraphics[height=3.65in,angle=0]{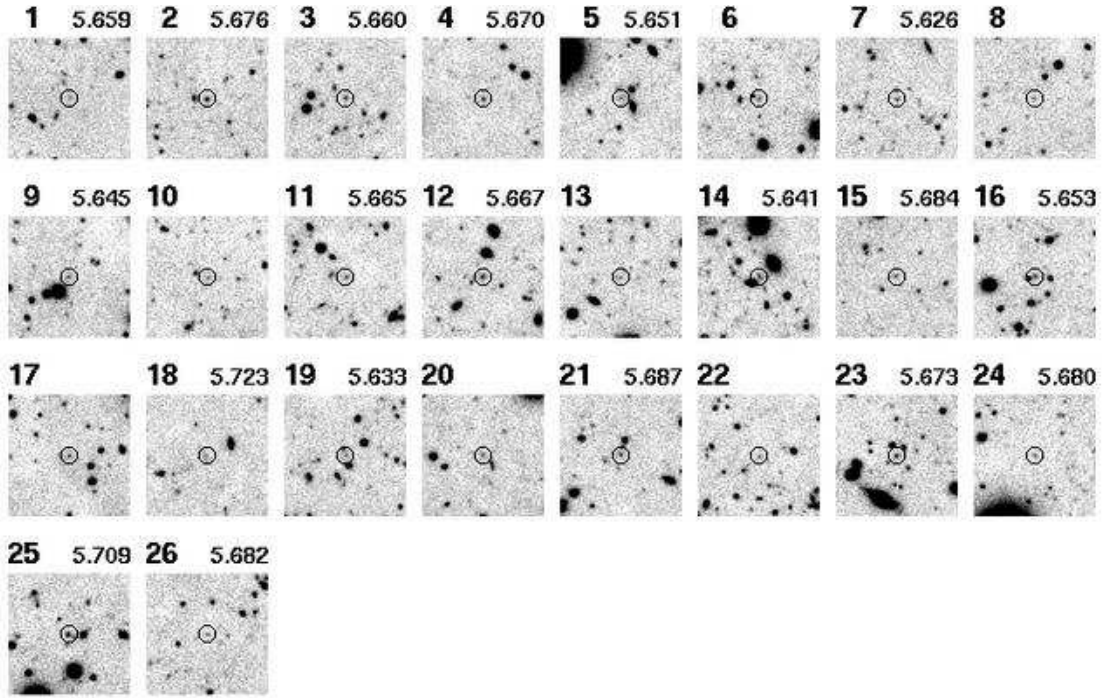}
\caption{Narrowband 8150/120 \AA\ images 
    of the 26 emission-line candidates selected in 
    \protect{Table~\ref{tbl-3}}. Each field is 30$''$ on a side. 
    A 2$''$ radius circle identifies each candidate, which is labeled
    by object number and redshift (for confirmed high-$z$ galaxies). 
    This figure may be contrasted
    with the corresponding $R$-band images of
    Fig.~\protect{\ref{fig8:r-thumbs}}. 
   \label{fig7:nb-thumbs}}
\vspace*{-0.4cm}
\end{figure*}

\begin{figure*}
\vspace*{-0.1cm}
\centering
\includegraphics[height=3.65in,angle=0]{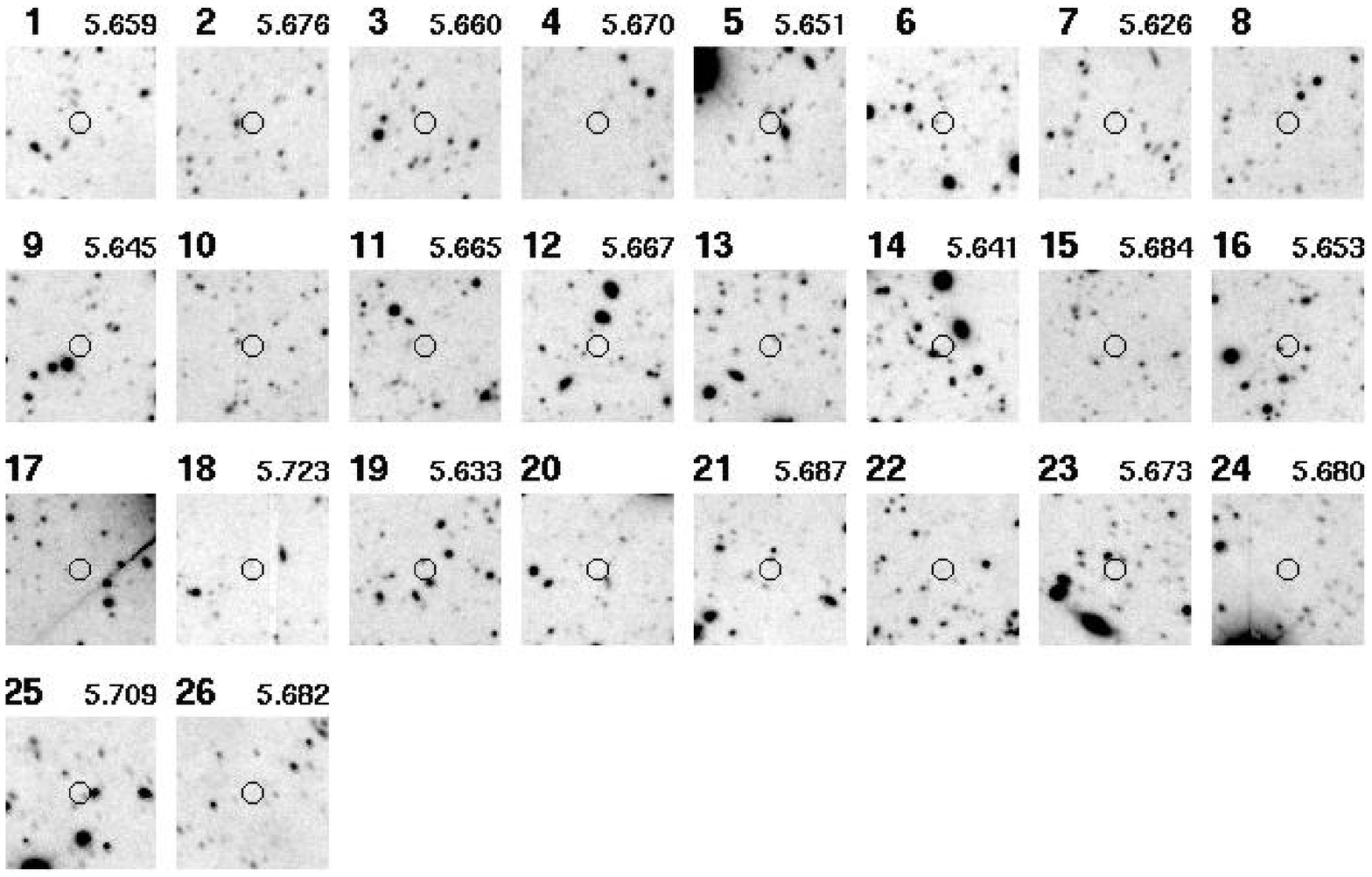}
\caption{$R$-band images of the selected emission-line candidates
   shown in 
   Fig.~\protect{\ref{fig7:nb-thumbs}}.  A 2$''$ radius circle
   indicates the candidate's location in each $30''\times30''$
   field. Candidates are labeled by object number and redshift
   (for confirmed high-$z$ galaxies)
   from \protect{Table~\ref{tbl-3}}. Absorption due to the Lyman
   alpha forest should suppress $R$-band light from redshift
   $z\sim5.7$ galaxies.
   \label{fig8:r-thumbs}}
\vspace*{-0.4cm}
\end{figure*}

This final candidate sample contains 26 objects.  Tabulated coordinates,
multi-color magnitudes, and redshifts (where measured) for these
objects are summarized in Table~\ref{tbl-3}.  Thumbnail finding
charts in the narrowband (Fig.~\ref{fig7:nb-thumbs}) and $R$-band
(Fig.~\ref{fig8:r-thumbs}) are shown, with the objects encircled.
Each thumbnail is 30$''$ on a side, with the circles $2''$ in radius.

\section{Spectroscopic classification }
\label{secclass}

Twenty-two of the 26 objects in the candidate list were observed with DEIMOS
with exposure times ranging from 1.5 to 4.5 hours. A randomly chosen
sample of 21 additional objects with $(I-N)>0.7$ which did not satisfy
the color selection was also observed with the choice of objects
in this case being determined by the mask configurations. One additional
object in the candidate list which had been previously observed with LRIS
was not included in the DEIMOS masks. The DEIMOS masks were filled
with color selected and magnitude selected samples which will be described
elsewhere and contained just over 900 objects.

All of the spectra (emission line and field objects) were spectroscopically
classified without reference to their narrow band strengths and color 
properties.  As can be seen from Fig.~\ref{fig2:composite-spec} at the 
DEIMOS resolution \oii\ and \oiii\ 
lines are easily distinguished based on their doublet structure, while
H$\alpha$ emitters can usually be picked out on the basis of the [N$\,$II]
or [S$\,$II] doublets or using the H$\beta$, [O$\,$III] complex. 
Only a small number of emission line objects show single broad asymmetric 
lines of the type illustrated in Fig.~\ref{fig2:composite-spec} and these 
were classified as Ly$\alpha$ emitters. Where there was uncertainty
in the classification the object was marked as observed but unidentified.
Where an object was observed multiple times each spectrum was measured
individually. However, there was no disagreement among identifications
and in the final compilation these spectra were averaged.

\begin{figure}
\vspace*{-0.6cm}
\includegraphics[width=2.8in,angle=90,scale=0.9]{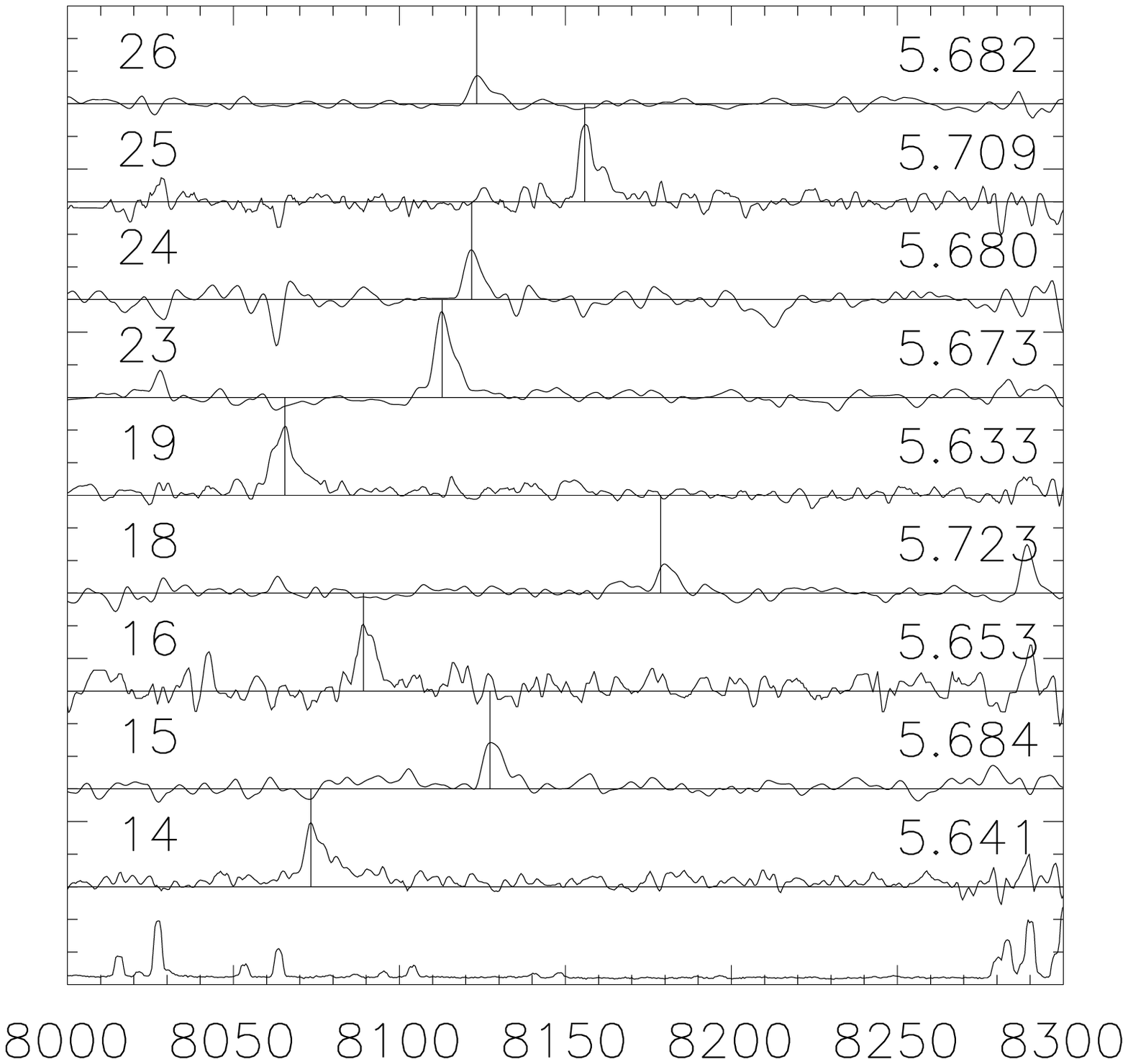}
\vspace*{-0.4cm}
\includegraphics[width=2.8in,angle=90,scale=0.9]{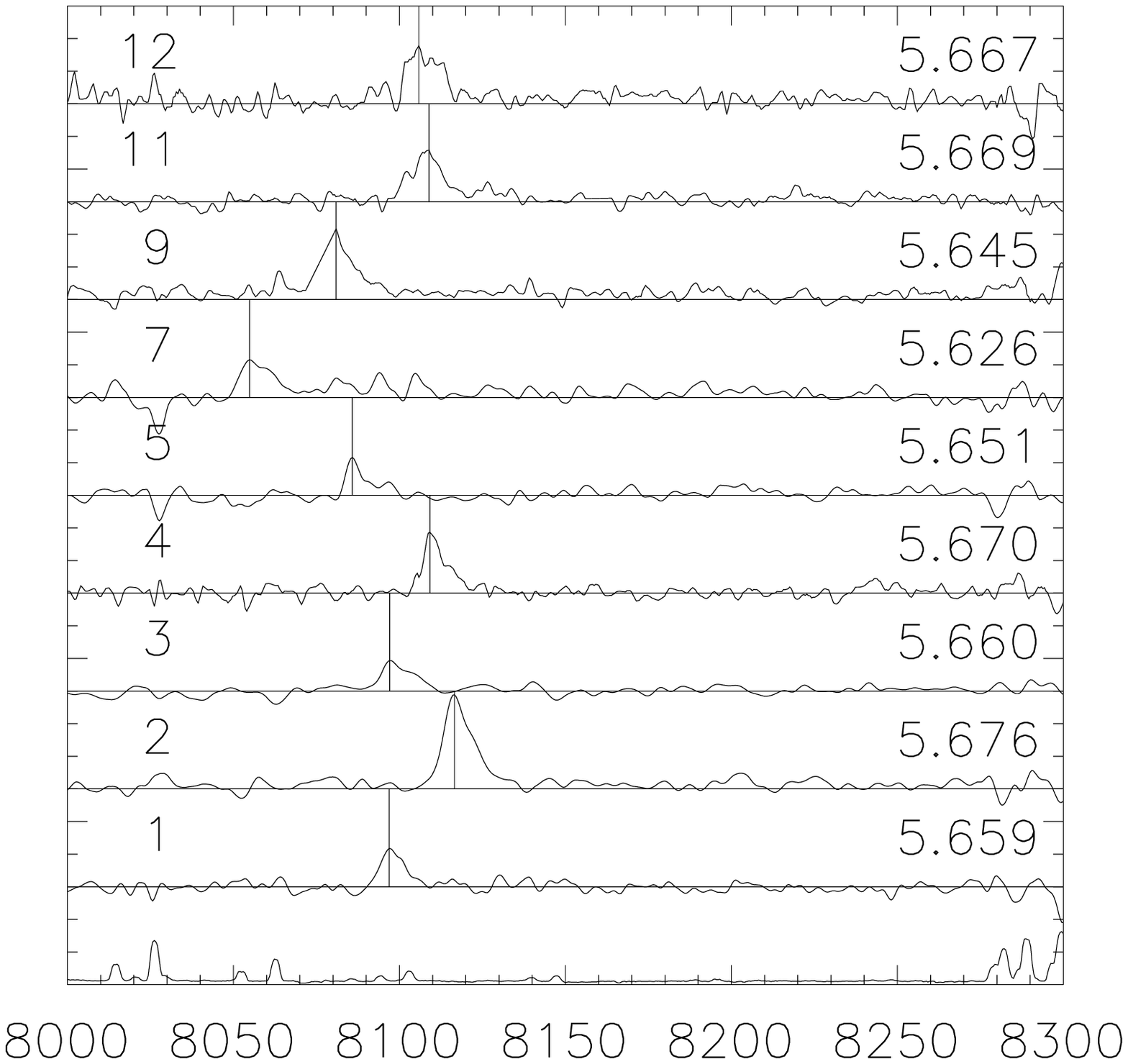}
\caption{Extracted DEIMOS spectra of single-line emission line objects 
   with asymmetric profiles in the candidate sample.  Spectra are labeled 
   with the assigned object number and redshift given in Table 3. 
   The bottommost plots show the positions of
   strong nightsky lines for comparison with subtraction residuals
   (e.g., objects 18, 23, 24, 25, 12, 4, and 2).  
  \label{fig9:la-spectra}
  } 
\vspace*{-0.2cm}
\end{figure}

\begin{figure}
\vspace*{-0.4cm} 
\includegraphics[width=2.8in,angle=90,scale=0.9]{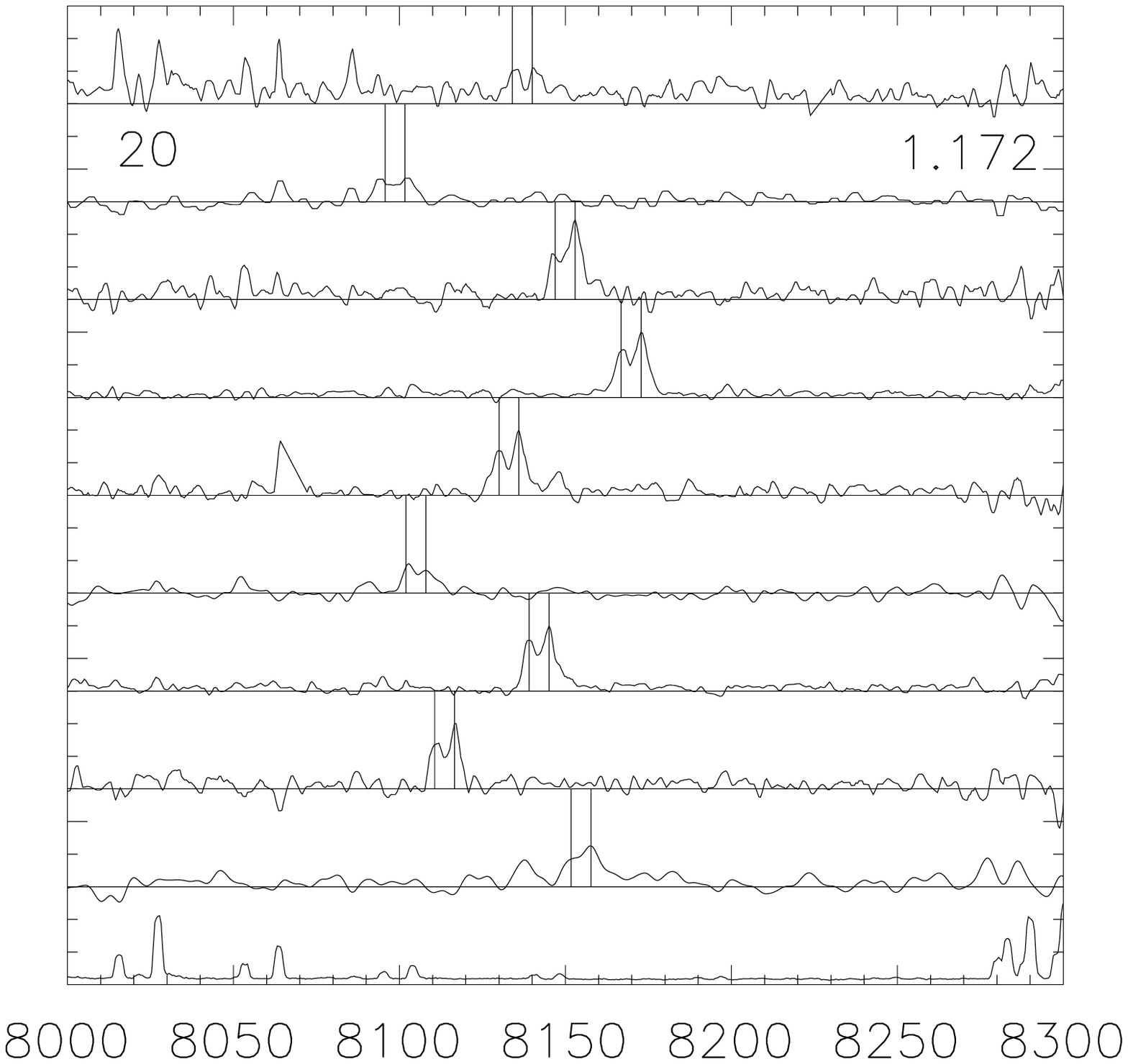}
\caption{Spectra of emission line objects with the [O$\,$II] doublet signature.
   Candidate object 20, from \protect{Table~\ref{tbl-3}} 
   is identified as an [O$\,$II] emitter
   at redshift $z=1.172$. The remaining 8 spectra show galaxies selected
   according to [O$\,$II]  emission excess within the narrowband filter.
   As for Fig.~\protect{\ref{fig9:la-spectra}},
   the bottom spectrum shows the positions of the strong
   nightsky lines which may give rise to subtraction residuals.
  \label{fig10:o2-spectra}
  } 
\end{figure}

\begin{figure}
\includegraphics[width=2.8in,angle=90,scale=0.9]{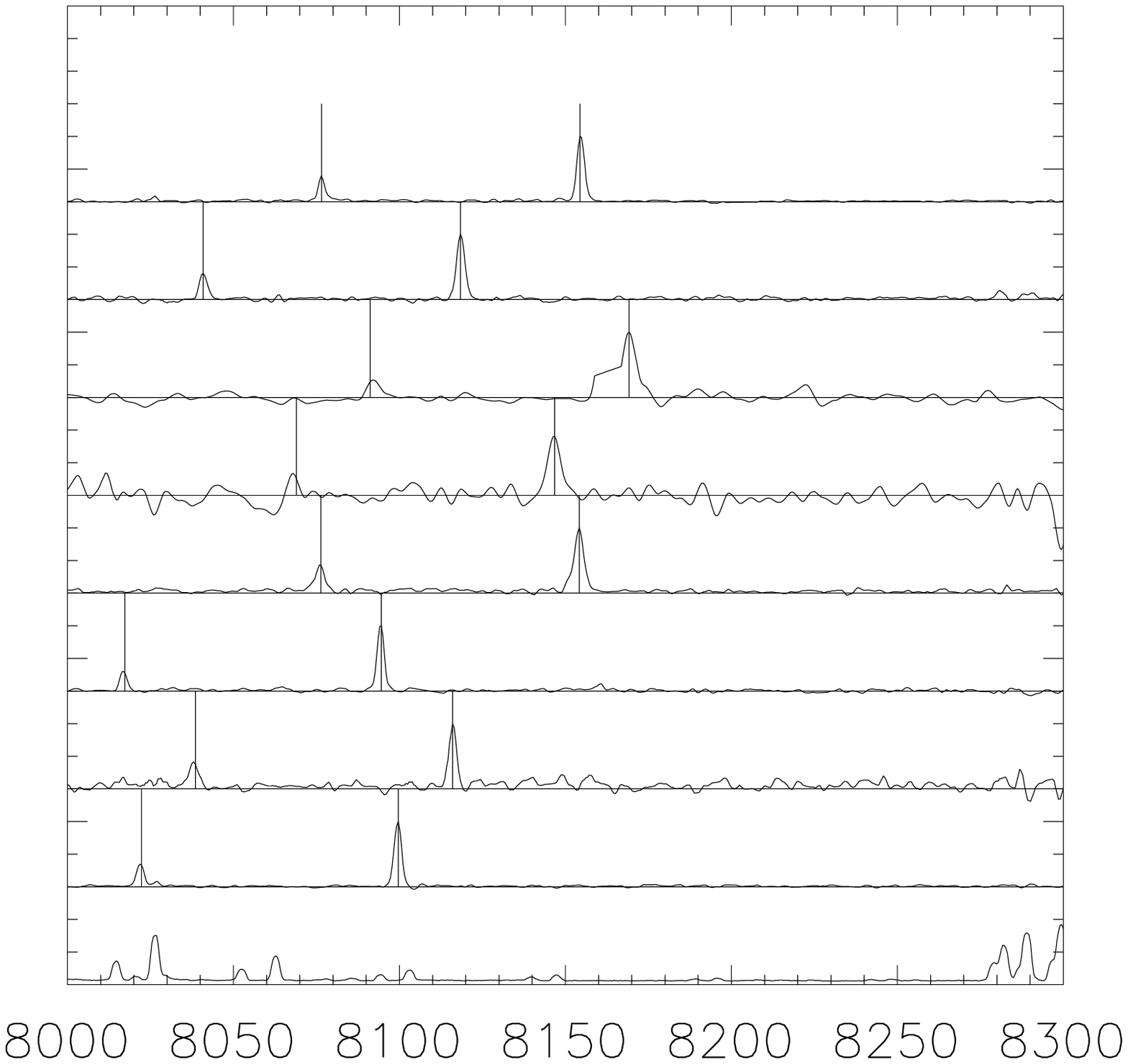}
\caption{Spectra of emission line objects with the [O$\,$III] doublet signature.
   As in Fig.~\protect{\ref{fig9:la-spectra}},
   the bottom spectrum shows the positions of the strong
   nightsky lines which may give rise to subtraction residuals.
  \label{fig11:o3-spectra}
  } 
\end{figure}

\begin{figure}
\includegraphics[width=2.8in,angle=90,scale=0.9]{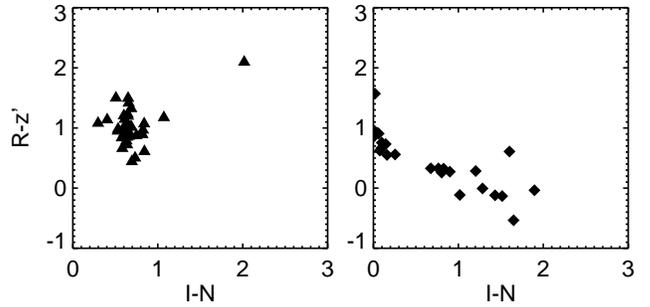}
\caption{Color-color plots in $R-z'$ vs.\ $I-N$ of [O$\,$II] emitters 
  (triangles) and [O$\,$III] emitters (diamonds).  [O$\,$II] emission-line
   galaxies display a shallow continuum break in ($R-z'$) of about 1
   magnitude.  The range of emission-line excess ($I-N$) reflects
   the cut-off of selection criteria at the low excess end, but
   is generally modest.  The [O$\,$III] emitters in the narrow-band
   filter can display stronger observed-frame excess emission,
   reflected in the wider range of emission excess in ($I-N$).  The
   [O$\,$III] galaxies have generally flat spectra ($R-z'$) $\sim 0$,
   but there is a bluing trend seen in the ($R-z'$) colors for
   galaxies with very strong [O$\,$III] emission lines. The one
   object classified as an [O$\,$II] emitter (object \#20)
   in the candidate sample
   lies in the upper right corner of the left hand panel.
  \label{fig12:o2_colors}
  }
\end{figure}

The redshift identifications of the high $z$ candidate list are summarized in
Table~\ref{tbl-3}. Of the 22 DEIMOS observed objects 18 were identified 
as Lyman alpha emitters, as was the LRIS observed object. One object was 
classified as an [O$\,$II] emitter
and three could not be clearly identified. The spectra of the identified 
$(N-I) > 0.7$ objects are
shown in Figures \ref{fig9:la-spectra},
\ref{fig10:o2-spectra}, and
\ref{fig11:o3-spectra}. One of the unidentified objects has a weak
continuum spectra which probably corresponds to a star and lies in the
area of $z'-N$, $I-N$ color-color space consistent with this interpretation
as does one of the unobserved objects (17 and 8 respectively). One object
(21) has a strong red emission line which is symmetric and narrow and which, 
if it is H$\alpha$, places the object at redshift $z=0.492$.  There is a 
weak feature at the [O$\,$III] $\lambda$5007 emission position which could
be consistent with this interpretation but, irrespective, this object is
clearly not a $z\sim5.7$ Lyman alpha emitter. Finally, a retrospective analysis
of the [O$\,$II] emission line classified object (20) suggests it is ambiguous
whether it should have been classified in this way since the [O$\,$II]
lines are slightly displaced from the correct positions. It is
possible this is actually a Lyman alpha emitter.  As we show in
Fig.~\ref{fig12:o2_colors} if this object is an [O$\,$II] emitter
its colors are extremely anomalous.

For the remainder of the work we will focus on the 19 spectroscopic
Ly$\alpha$ emitters recognizing that a further 4 objects could fall
in the sample. This small level of incompleteness does not affect the discussion
in any statistically significant way.

\section{Results and Discussion}
\label{secdisc}
\subsection{Spatial and spectral distributions}
\label{secdist}
\begin{figure}
\includegraphics[width=2.8in,angle=90,scale=0.9]{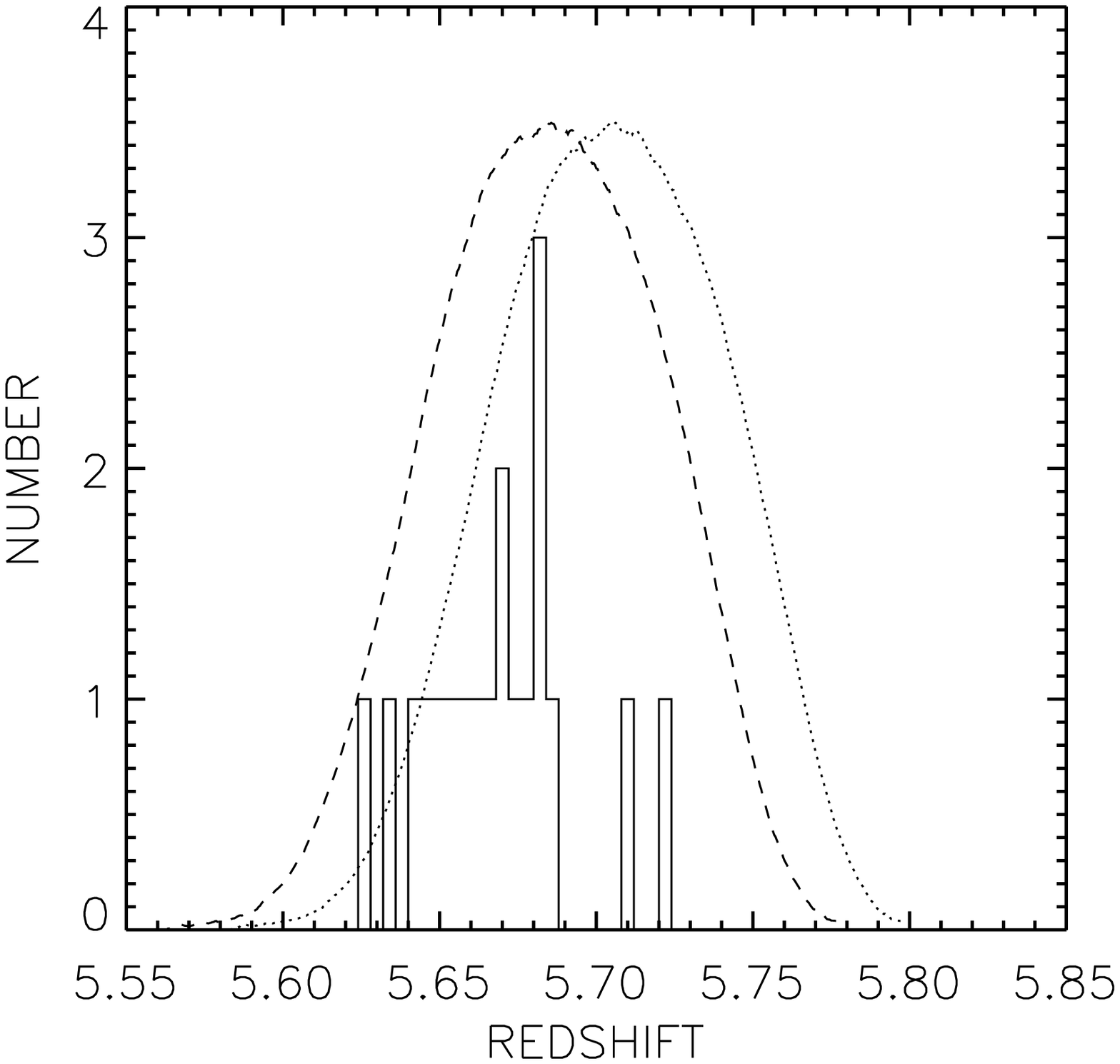}
\caption{The redshift distribution of Ly$\alpha$ emission galaxies 
   compared to the transmission profile of the narrow band filter
   (dotted line). Nearly all of the objects lie at the low
   redshift end of the selection, This result remains
   true even if we shift the filter
   center $25$~\AA\ to the blue (dashed line).
  \label{fig13:zdist}
  }
\end{figure}

In Figure~\ref{fig13:zdist} we show the distribution of the 19 Lyman alpha redshifts
compared with the nominal transmission profile of the filter
shown by the dotted line. All but one of the 19 objects lie below
the redshift corresponding to the $8150 \AA$ central wavelength of the
filter. The [O$\,$III] emitters also lie in the lower half of the filter
and taken together these results could suggest that the true on-telescope
center of the filter is shorter than the measured value. However, both
could be the result of structuring at the respective redshifts.
Even if we shift the filter center 25 \AA\ blueward to match the
[O$\,$III] data (dashed line) 16 of the 19 Ly$\alpha$ emitters still lie in the
lower half of the filter. The probability of seeing such a biased
result is only 0.0003 while with the laboratory traced filter profile
the probability would drop to $2\times10^{-6}$. It appears that the
sample is extremely structured in velocity space.

\begin{figure}
\includegraphics[width=2.8in,angle=90,scale=0.9]{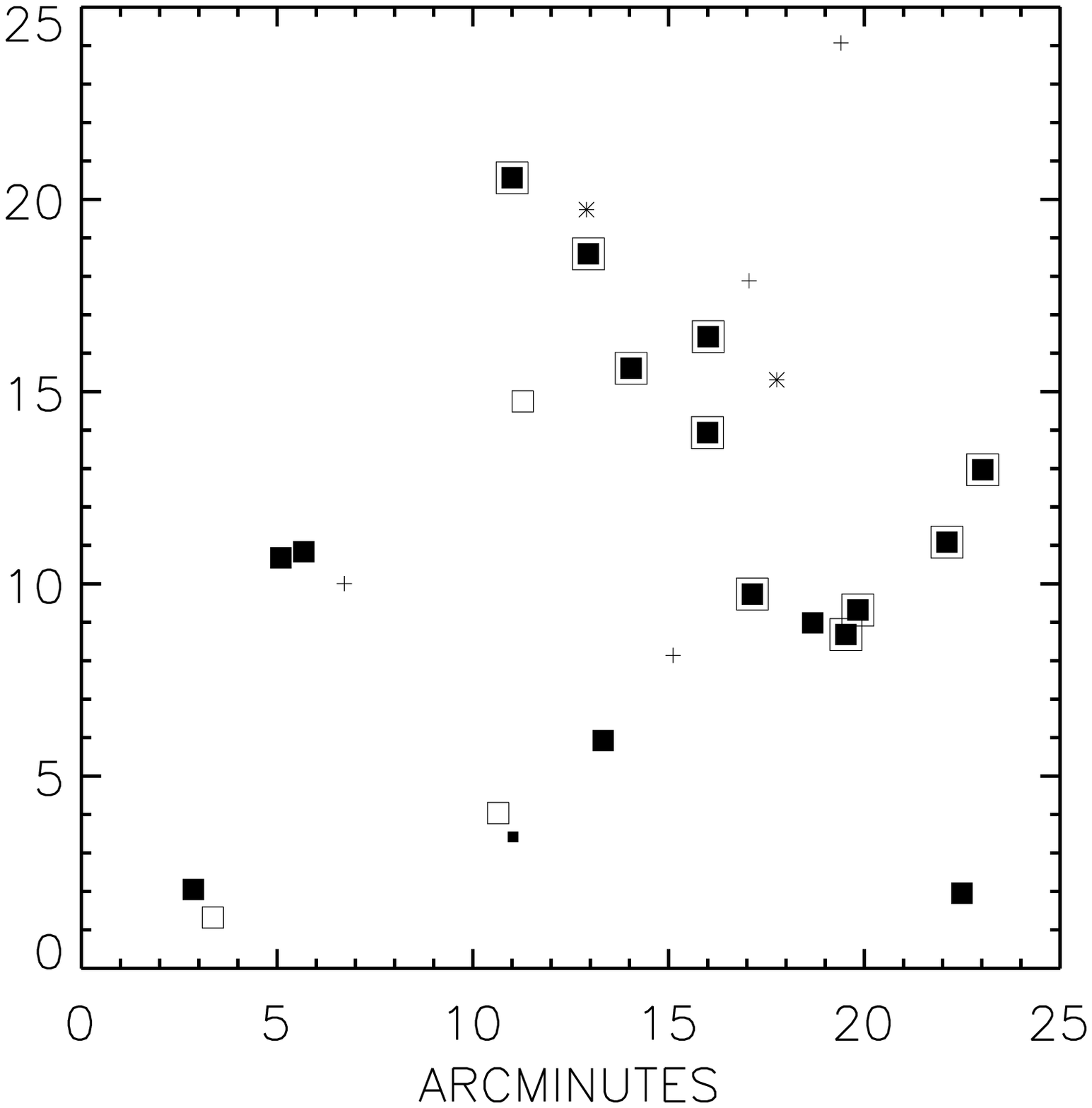}
\caption{The spatial distribution of the 19 Ly$\alpha$ emission galaxies 
   is shown by the large solid ($z < 5.687$) and open ($z >5.687$) boxes.
   The remaining objects in the candidate list are shown as crosses
   for unidentified objects, asterisks for possible stars, and the
   small box for the possibly misclassified [O$\,$II] emitter which, if
   it were a Lyman alpha emitter, would lie in the lower redshift
   interval. Many of the low $z$ objects lie in the large diagonal
   filament which runs across the field. Most of the objects
   in this region lie in the more restricted redshift interval
   $z=5.632-5.672$. Objects lying in this redshift region are
   enclosed in larger open boxes.
   \label{fig14:structure}
  }
\end{figure}

The objects lying in the lower redshift interval also
appear to be clustered in space, with many of them
lying in a long elongated filament running diagonally
across the field (Figure~\ref{fig14:structure}). All of the objects
between $z=5.632-5.672$ lie in this region. 

We postpone a more detailed discussion of the correlation
function properties until the analysis of the multi field
sample but the present results emphasize that single fields
may vary considerably even with areas as large as the present
ones and that many independent fields are required to average
out these systematic effects.

\subsection{Line Shapes}
\label{secprof}
\begin{figure}
\centerline{\epsfig{file=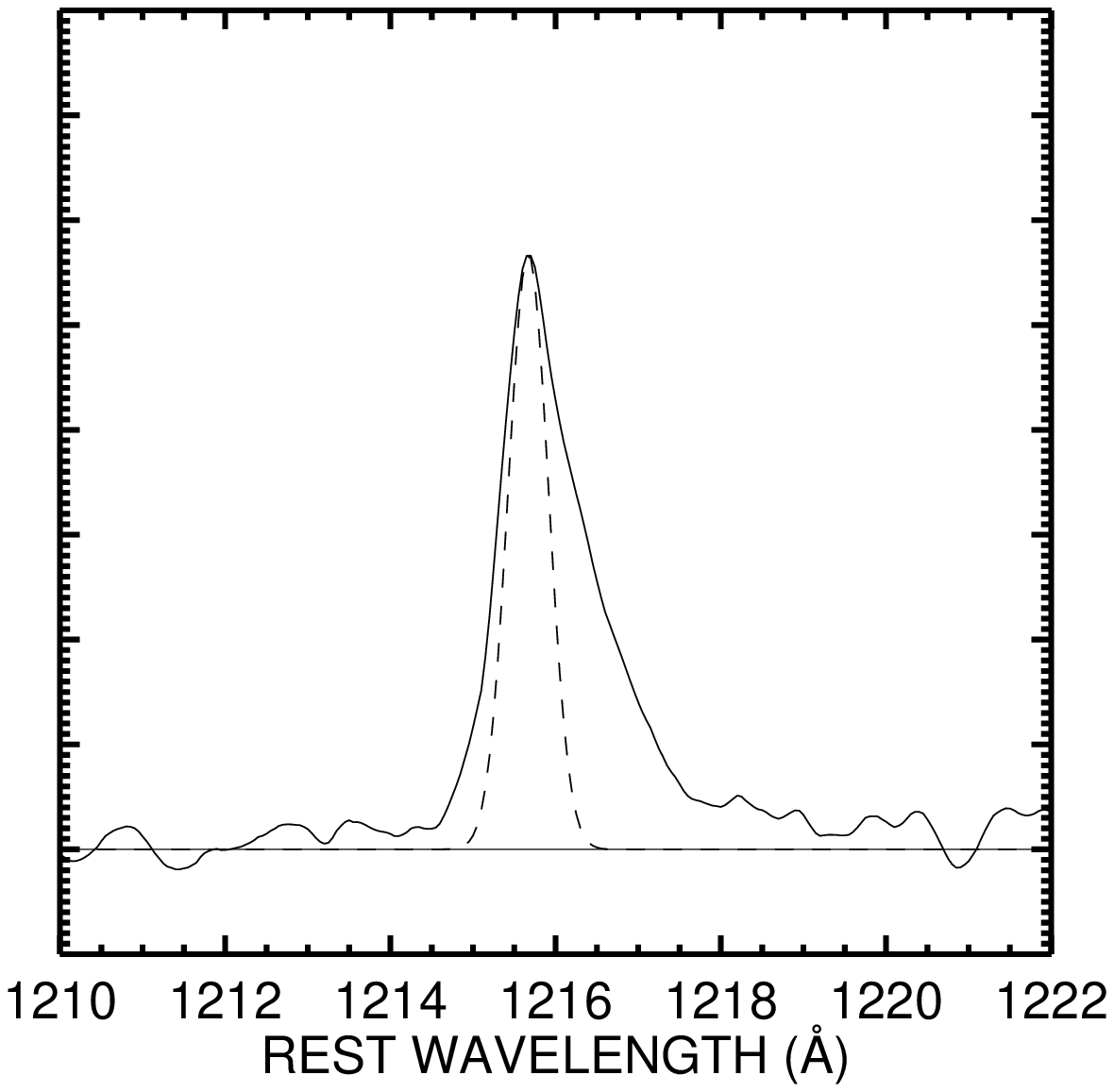,width=2.8in,angle=0,scale=1.0}}
\caption{The stacked Ly$\alpha$ emission line profile of the eighteen
   new $z\sim5.7$ galaxies reported in this paper.  Individual 
   extracted spectra were normalized to their maximum value, converted
   to their rest-frame wavelength scale using the redshift corresponding
   to their peak flux, and co-added.  The instrumental profile as determined
   from neighboring resolved nightsky lines is overplotted on the peak
   for comparison.  The asymmetric profile can be clearly seen.
  \label{fig15:la-stack}
  }
\end{figure}

\begin{figure}
\includegraphics[width=2.8in,angle=90,scale=0.9]{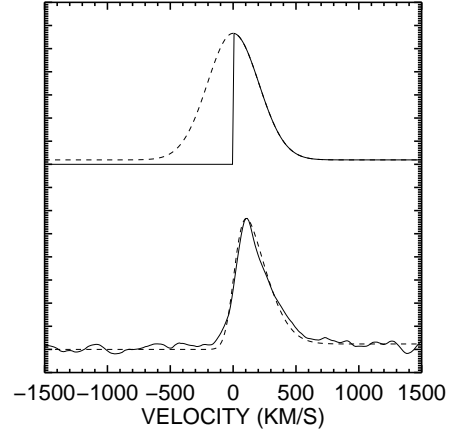}
\caption{A simple model profile for the Ly$\alpha$ emission feature
   in Fig.~\protect{\ref{fig15:la-stack}}.  The top profile shows a 
   Gaussian emission-line
   profile with Doppler width $\sigma=200$~km~s$^{-1}$ with the
   blue half of the line (dashed curve) absorbed by the foreground Ly$\alpha$
   forest.  When this is convolved with the instrumental profile
   of the spectrograph, the resulting profile (dashed curve) shows a good
   fit to the observed stacked Ly$\alpha$ profile (solid curve) in the
   bottom half of the figure.
  \label{fig16:z5-model}
  }
\end{figure}

As can be seen from Figure~\ref{fig9:la-spectra} the Ly$\alpha$ 
shapes are remarkably
uniform, consisting of a slightly peaked profile and a trailing
red wing. In Figure~\ref{fig15:la-stack} we show the stacked profiles of the 18
Ly$\alpha$ emitters with DEIMOS spectra 
compared with the resolution element of the instrument, which clearly
illustrates this point, but individual profiles are also generally similar
in shape and width. This type of profile is a simple outcome
of the blue truncation of the spectrum as is illustrated in 
Figure~\ref{fig16:z5-model} where we show a simple model where the intrinsic
line is modeled as a Gaussian and then is truncated below
the object redshift by the Lyman alpha scattering of the
IGM.  When this profile is convolved through the instrument
resolution it produces an observed spectrum almost identical
to that observed as is shown in the lower panel of the figure.

The required width of the intrinsic line which matches
the observed stacked spectrum corresponds to a $\sigma$
of 200 km~s$^{-1}$ and this value also produces a reasonable fit
to nearly all the individual line profiles. This is much
wider than would be expected from the velocity dispersions
of the parent galaxies at this redshift and almost certainly represents
velocity broadening of the Ly$\alpha$ during the escape
from the galaxy. Given the complexity of this process
there is probably relatively little useful information
in the widths. The truncation also results in the
measured redshifts, which were set to the peak of the
line, being about a 100 km~s$^{-1}$ redward of the true redshift,

\subsection{Equivalent Widths}
\label{secew}
Since the narrow band lies almost precisely in the center
of the $I$ band the truncation effect reduces the $I$ band
flux and the narrow band flux in very similar amounts
and we shall assume that the equivalent widths (EWs) of the
lines do not need to be corrected for this effect. 
(This approximation does depend on the IGM scattering
not extending substantially redward of the line center
which could be incorrect.)  Because the filter is
not rectangular in shape
we do need to allow for the filter profile in computing
the EWs.  

The simplest approach is to compute the equivalent
widths based on the $I$ and $N$ magnitudes. Defining
the quantity 
\begin{displaymath}
	R=10^{-0.4*(N-I)}
        \label{eq:3}
\end{displaymath}
the observed frame equivalent width becomes
\begin{displaymath}
        EW = \left [ {R - 1} \over {\displaystyle {\phi - 
             {R\over{\Delta\lambda}}}} \right ]
	\label{eq:2}
\end{displaymath}
where $\phi$ is the narrow band filter response normalized
such that the integral over wavelength is unity and $\Delta\lambda$
is the effective width of the $I$ filter, which is approximately
1400 \AA. The narrow band filter is often assumed to
be rectangular in which case $\phi$ becomes $1/\Delta\lambda$
where $\Delta\lambda$ is the width of the narrow band
\citep[e.g.,][]{rho02} but as can be seen
from Figure \ref{fig13:zdist} this is not a very good approximation
in the present case.

For very high equivalent width objects such as the Lyman alpha
emitters the denominator
in this equation becomes uncertain unless the broad band data is
very deep and this can result in a very large scatter in the measured
equivalent widths. Therefore in order to check the result we
also computed the equivalent widths by extrapolating from the
$z'$ observations assuming that these provide a better measure
of the true continuum flux. We assumed in this case that the
spectral energy distribution was flat in frequency space
giving the formula 
\begin{displaymath}
        EW = \left [{2 \times {\rm Norm} \times (R_2 - 1)} \over {\displaystyle
          {\phi}} \right ]
        \label{eq:4}
\end{displaymath}
where 
\begin{displaymath}
	R_2 = 10^{-0.4*(z'-N)}
	\label{eq:5}
\end{displaymath}
and Norm is the square of
the wavelength ratio of the center of the $z'$
filter over that of the narrow band.
In order to avoid uncertainties from
the exact wavelength center of the filter we restrict
ourselves to wavelengths where the nominal filter response
is more than 0.67 of the peak. 

\begin{figure}
\includegraphics[width=2.8in,angle=90,scale=0.9]{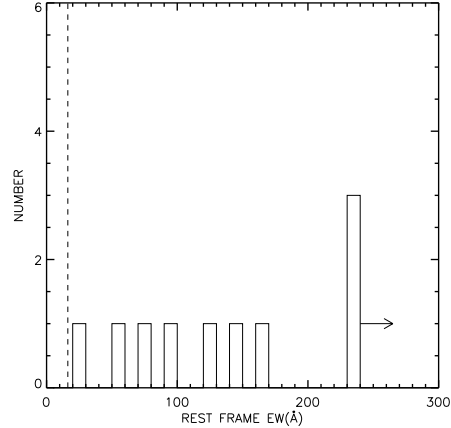}
\caption{Rest frame equivalent width distribution for the 15
Lyman alpha emitters lying in the redshift range where the
filter profile lies above 0.67 times its maximum value.
  \label{fig17:ew-dist}
  }
\end{figure}

The rest frame equivalent widths calculated with both methods
generally agree well with each
other though there are significant differences in a small
number of individual cases. The distribution based on the
extrapolated fluxes is shown in Fig.~\ref{fig17:ew-dist}. In contrast
to the results of \citet{mal02} most of the
rest frame equivalent widths have values less than
240~\AA, which can be understood in terms of galaxies
with Salpeter initial mass functions. (See Malhotra
and Rhoads for an extensive discussion of this issue.)
As discussed in the previous paragraph, the equivalent
width determination is very sensitive to the depth of
the continuum band exposures, and the difference between the 
present results and those of \citet{mal02} may arise
from the deeper continuum exposures used here.
Only four of the objects have poorly defined rest
frame equivalent widths and are shown at nominal values
of 240~\AA\ in the figure but even these objects could
well lie at lower values based on the uncertainties
in the determination. The overall distribution of
Fig.~\ref{fig17:ew-dist} is quite similar to the distributions measured
at lower redshifts (e.g. Figure 21 of \citet{fuj03}
which summarizes these measurements.)

\begin{figure}
  \includegraphics[width=2.8in,angle=90,scale=0.9]{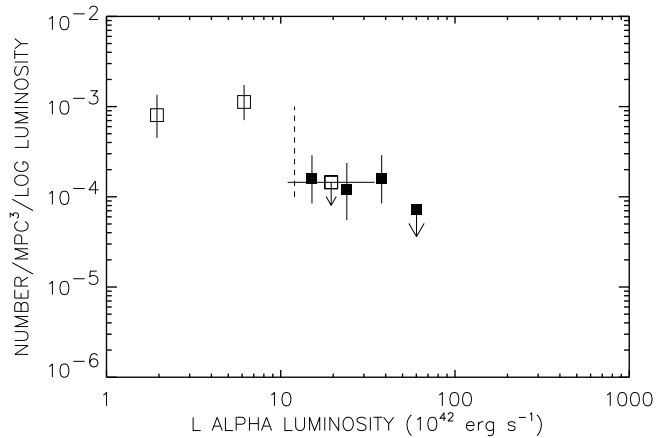}
  \caption{The luminosity function based on the Ly$\alpha$ emission line
    for the $z\sim5.7$ emitters (filled squares) compared to the
    emission line luminosity function at $z\sim3.4$ (open squares).
    The vertical dashed line indicates the limits of the sample,
    which does not yet reach the flat part of the $z\sim5.7$ luminosity
    function.  
  \label{fig18:emlum-funct}
}
\end{figure}

\subsection{Lyman alpha Luminosity Function}
\label{seclal}

Because of the high observed frame equivalent widths the
Ly$\alpha$ fluxes are insensitive to the continuum determination.
However, they do depend on the filter response at the emission
line wavelength so we have again restricted ourselves to
redshifts where the nominal filter response is greater than
67\% of the peak value. 
 
Because the volume is simply defined by the selected redshift
range the luminosity function may be obtained by dividing the
number of objects in each luminosity bin by the volume.
The calculated Ly$\alpha$ luminosity function  is shown
in Figure~\ref{fig18:emlum-funct}, where we compare it with the $z=3$ luminosity
function computed from the results of \citet{smitty}.  The observations
of \citet{kud00} give a similar value at this redshift. The present function 
samples generally higher luminosities than the $z=3$ samples
but is broadly consistent in number density at the lowest
luminosities. It lies slightly above the \citet{fuj03}
values at similar luminosities, but this probably reflects
the higher equivalent width selection that was used in that
work.

\begin{figure}
  \includegraphics[width=2.8in,angle=90,scale=0.9]{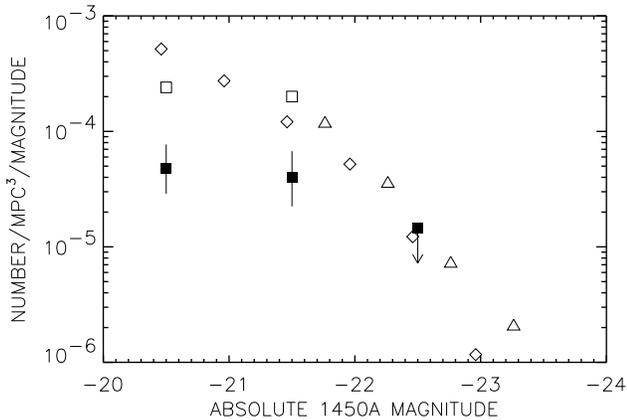}
  \caption{The luminosity function of the UV continuum of $z\sim5.7$ emitters
    compared with the Steidel et al. 1999 UV luminosity functions at
    redshifts 3 (diamonds) and 4 (triangles).  Measured points from the
    Ly$\alpha$ emitters are shown as solid boxes; the open boxes show
    assumed values if emitters pick out 20\% of the LBG sample as
    Steidel et al.\ (2000) find to be the case at $z\sim3$.
  \label{fig19:contlum-funct}
}
\end{figure}

\subsection{Continuum Luminosity Function}
\label{seccont}

We may also view the Lyman alpha selection as picking out that subset
of the Lyman break galaxies (LBGs) at this redshifts which have strong line emission.
While some of the objects are very faint in the $z'$ band, which corresponds
to a rest frame of about $1400$~\AA, most (16 of the 19) have well measured $z'$
magnitudes lying in the range from 24 to 27. These magnitudes correspond
to absolute continuum magnitudes at $1450$~\AA\ which are very similar
to the typical LBG at $z=3-4$. 
The use of continuum magnitudes to estimate the star formation rates
avoids the extremely complex problems of the Lyman alpha escape process
and the uncertainties in the correction of the Lyman alpha fluxes
for intergalactic scattering which are present in the determination of the 
Ly$\alpha$ luminosity function.

Figure \ref{fig19:contlum-funct} shows the continuum luminosity
function derived from the distribution of the $z'$ magnitudes. Once
again we have restricted ourselves to objects lying in the higher
transmission region in the filter. The solid boxes show the measured
values with 1 $\sigma$ Poisson uncertainties based on the number of
objects in the bin. There are no objects with absolute magnitudes
brighter than $-22$ and for the final data point we show a 1 $\sigma$ upper
limit. No correction has been applied for dust. Figure \ref{fig19:contlum-funct} 
also shows the raw (dust uncorrected) UV luminosity function for Lyman break selected galaxies
at redshifts $z\sim3$ (open diamonds) and $z\sim4$ (open triangles)
\citep{stei99,stei00}. At these redshifts \citet{stei00} found 
the Lyman alpha selection picks out about $20\%$ of the Lyman break galaxies.
The open squares show the values which would be obtained for the LBG
luminosity function at $z=5.7$ if the Lyman alpha selected fraction were similar
to that at the lower redshift. It matches extremely closely to the
lower redshift luminosity functions.  The Ly$\alpha$ selected fraction at 
these redshifts could be larger, and comparison with the color-selected
samples of \citet{stan03a} would be consistent with this interpretation.

Once again, recognizing the systemic uncertainties in a single field
observation and the uncertainties involved in the Ly$\alpha$ selected fraction, we postpone a more detailed discussion to a subsequent
paper. However, it is clear already from Figure \ref{fig19:contlum-funct}
that star formation rates at $z=5.7$ are roughly similar to those
at lower redshift and there is no rapid decline in the star formation
rates at these redshifts.

\section{Conclusions}
\label{secconc}

We may summarize the major conclusions of the paper as follows.
\smallskip

1) Combined emission line and color break techniques provide an
extremely efficient method of selecting objects at $z=5.7$.
Selection criteria of $(N-I) > 0.7$, $(R-z') > 1.8$ and $(N-z') < 0.2$
pick out 24 objects in the field. At least 19 of these are
spectroscopically confirmed as Lyman alpha emitters at this
redshift. Only one or at most two of the sample selected in this
way are lower redshift interlopers. The accuracy is sufficiently
good to allow use of color selected samples without the
necessity of spectroscopic followup for many purposes.

2) Even on scales as large as the present 26$'$ field the distribution
of objects is highly structured in both the spectral and spatial
dimensions. Individual fields could easily show factors of
several variation in the number density of objects selected
in a 100~\AA\ bandpass and a substantial number of fields need
to be observed to average out these effects.

3) The properties of the Lyman alpha selected galaxies are 
quite similar to those observed at $z \approx 3$.

4) Star formation rates at $z=5.7$ are roughly similar to those
at $z=3$ and $z=4$ though there are very substantial uncertainties
in this estimate.

\acknowledgements
This work was supported in part by the State of Hawaii and by NSF grants
AST-0071208 and AST99-84816, and by
NASA grant GO-7266.01-96A from Space Telescope Science 
Institute, which is operated by AURA, Inc., under NASA contract NAS 5-26555.  
We are grateful to the staff of the Subaru, Keck, and Canada-France-Hawaii
Telescopes for their support of these observations. We thank Sandy Faber,
Drew Phillips, Bob Kibrick, Alison Coil, Marc Davis and Grant Hill, and most 
especially, Greg Wirth, for their advice and assistance on DEIMOS, and
Sadanori Okamura and Masami Ouchi for their helpful inputs on the SuprimeCam
observations.

\end{document}